\documentclass[12pt,a4paper,prb,showpacs,preprint]{revtex4} \def\narrowtext{}\tighten

\usepackage[light]{draftcopy}
\usepackage{epsfig}
\newcommand{\beq}{\begin{eqnarray}}
\newcommand{\eeq}{\end{eqnarray}}  

\begin{document}
%%%%%%%%%%%%%%%%%%%%%%%%%%%%%%%%%%%%%%%%%
%\textwidth= 6.2 in
\textheight= 8.0 in
\topmargin= 0.25 in
\rightmargin= 1 in
\leftmargin= 1 in
\baselineskip=10 pt
\baselineskip=2\baselineskip
%%%%%%%%%%%%%%%%%%%%%%%%%%%%%%%%%%%%%%%%%
\vskip 3cm

\title{Characterization of ZnO:Si Nanocomposite Films Grown by 
Thermal Evaporation.}

\author{Shabnam Siddiqui}
 \affiliation{Department of Applied Sciences, 
Indira Gandhi Institute of Technology, 
Guru Gobind Singh Indraprastha University, Delhi 110 006, India.}
  \email{chhaya_rkant@yahoo.co.in}
\author{Chhaya Ravi Kant}
 \affiliation{Department of Applied Sciences, 
Indira Gandhi Institute of Technology, 
Guru Gobind Singh Indraprastha University, Delhi 110 006, India.}
  \email{chhaya_rkant@yahoo.co.in}
\author{P. Arun}
 \affiliation{Department of Physics \& Electronics, S.G.T.B. Khalsa College,
University of Delhi, Delhi - 110 007, India}
  \email{arunp92@physics.du.ac.in, arunp92@yahoo.co.in }
\author{N.C.Mehra}
 \affiliation{University Science Instrumentation Centre, University of Delhi, 
Delhi 110 007, India.}

\vskip 1cm

\begin{abstract}
Nanocomposite thin films of Zinc Oxide and Silicon were grown by
co-evaporating powdered ZnO and Si. This resulted in nanocrystallites of 
ZnO being embedded in Silicon. The mismatch in crystal structures
of constituent materials result in the ZnO nanocrystals to exist in a state
of stress. This along with oxygen vacancies in the samples result in 
good Photoluminescence emission at 520nm. 
Also, Silicon background gave a photoluminescence emission at 
620nm. The structure was found quite
stable over time since the homgenously dispersed ZnO nanocrystals do not 
agglomerate. The nanocomposites promises to be a useful candidate for future
optoelectronic devices.

\end{abstract}

\pacs{81.07.Bc, 81.07.Ta, 78.67.Hc, 07.79.Lh, 78.55.-m}

\maketitle
\narrowtext
\section{Introduction}
Since the fabrication of blue wavelength LED in the last decade,
research has gone on to attain LEDs of all possible wavelengths, even in the 
Ultra Violet wavelength range. Nanotechnology presents a possible method of
achieving this objectivity. It has been observed that in the nanometer regime, 
the band gap of a semiconducting material increases with reduction of its
particle/ grain size.\cite{nano} This presents a possibility of manufacturing materials 
of any desirable optical and electronic properties. 
However a problem confronted is that materials existing as nanoparticles 
tend to agglomerate, giving particles of larger size.\cite{nano2} To 
prevent further growth of particles capping agents were added during 
fabrication. This method has been extensively used in liquid phase. 
Alternatively, especially in the solid phase, nanoparticles have been grown 
embedded in a featureless background which is unmarked as far as its
properties are concerned in region of interest for that of the
nanoparticles.

\par Researchers working to develop LEDs/lasers which emit light with low 
wavelengths show immense interest in ZnO since it exhibits quantum 
confinement effects in the experimentally accessible range of sizes. Also,
it helps that ZnO is a wide band gap semiconductor materials, with 
possible application not only in electro-optical devices but also as 
varistors and transparent conducting films. ZnO nanoparticles also exist in
various morphological states like nanowires, nanorods, nanospheres
etc.\cite{r10,r11,r12,r13,r14,r15,r16,r17,r18,r19,r20,r21,r22,r23} 
This presents rich information from a pure science point of view for
understanding the basic quantum mechanical behaviour of matter in nanostate. 

\par Various processing techniques have been tried and developed to
manufacture nano-sized samples in a controlled fashion. Of the many methods,
a recent attempt by Singh et al \cite{forun} attracted our attention, where
the size of nanocrystalline ZnO was controlled by growing them on
porous-silicon substrates, whose pore size was controllable. Though their 
results were interesting, the fabrication process was involved
and in our opinion not commerically viable. We decided to investigate a
possible process where nanocomposites of zinc oxide and silicon could be
fabricated without the tedious process reported \cite{forun} yet present
features similar to those reported in the above mentioned study. In this 
article we report 
our observations on nanocomposites obtained by thermal co-evaporation of
silicon and zinc oxide.  

\section{Experimental Details}

ZnO:Si naocomposite films were grown by thermal evaporation method using 
JEOL, JEE 4X vacuum evaporation unit. n-Si wafers were ground into powder
and to this powder ZnO was added in a proportion of 2:1. The two were mixed
by grinding in an agat. Powdered ZnO of high purity (99.99\%) was obtained 
from Merck (Mumbai). Such well mixed material was then palletized to be used 
as the
starting material. The nanocomposite films were grown on microscopic glass 
substrates held at room temperature. The films were grown at vacuum better 
than ${\rm 10^{-6}}$-${\rm10^{-7}}$Torr. The structural studies of the films 
were done using X-ray diffractometer (Philips PW 3020). The film's surface 
morphology and texture were studied using the Scanning Electron Microscope 
(SEM) JEOL (JSM)-840 Scanning Microscope. Raman Spectra and
Photoluminescence were recorded using Renishaw's ``{\it Invia Reflex}'' Raman 
Spectroscope and Shimadzu's PL Spectroscope respectively.

%%%%%%%%%%%%%%%%%%%%%%%%%%%%%%%%%%%%%%%%%%%%%%%%%%%%%%%%%%%%%%%%%%%%%%%%%%%%%%%%%
\begin{figure}[h]
\epsfig{file=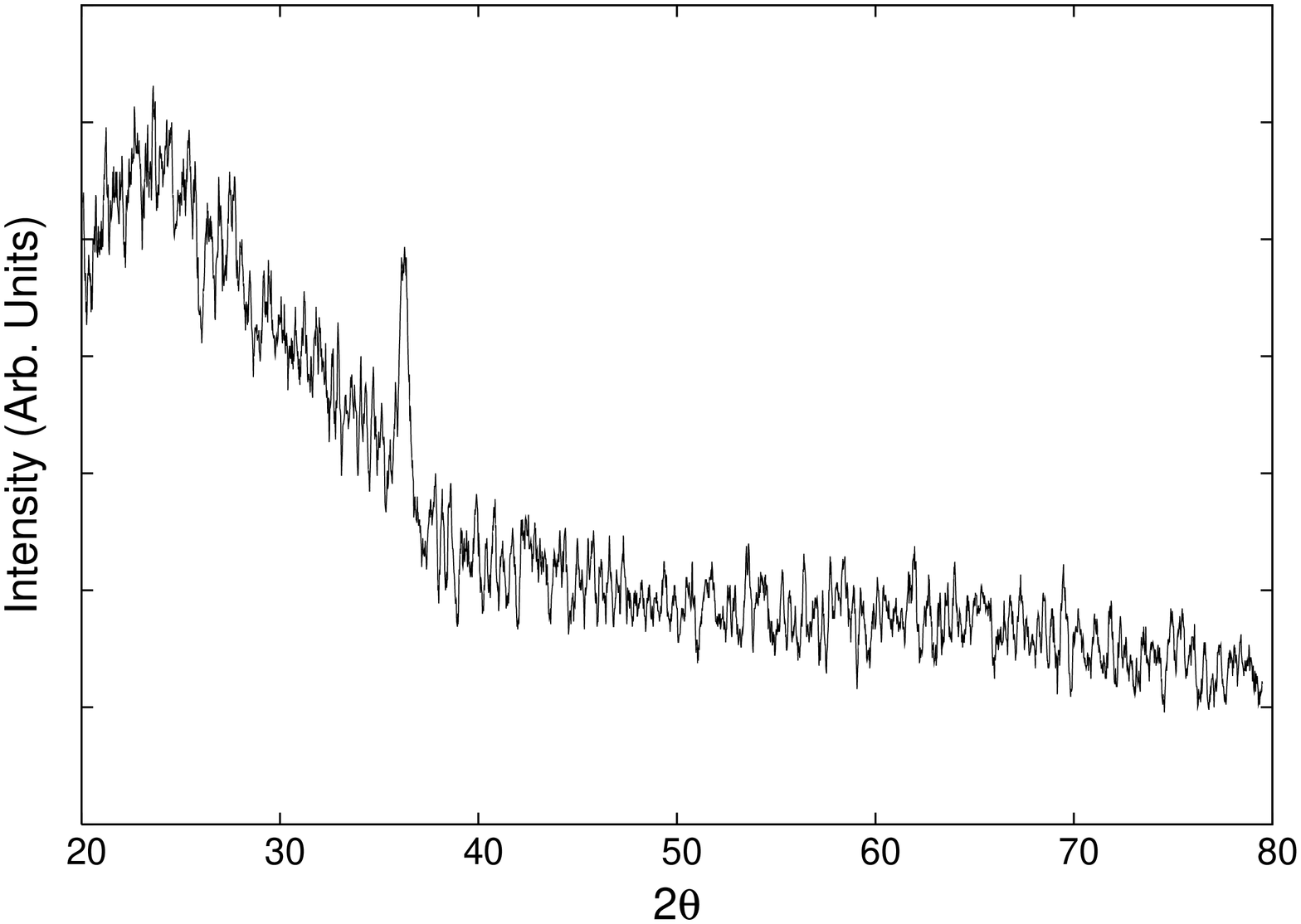, width=2in, angle=-90}
\vskip .4cm
\epsfig{file=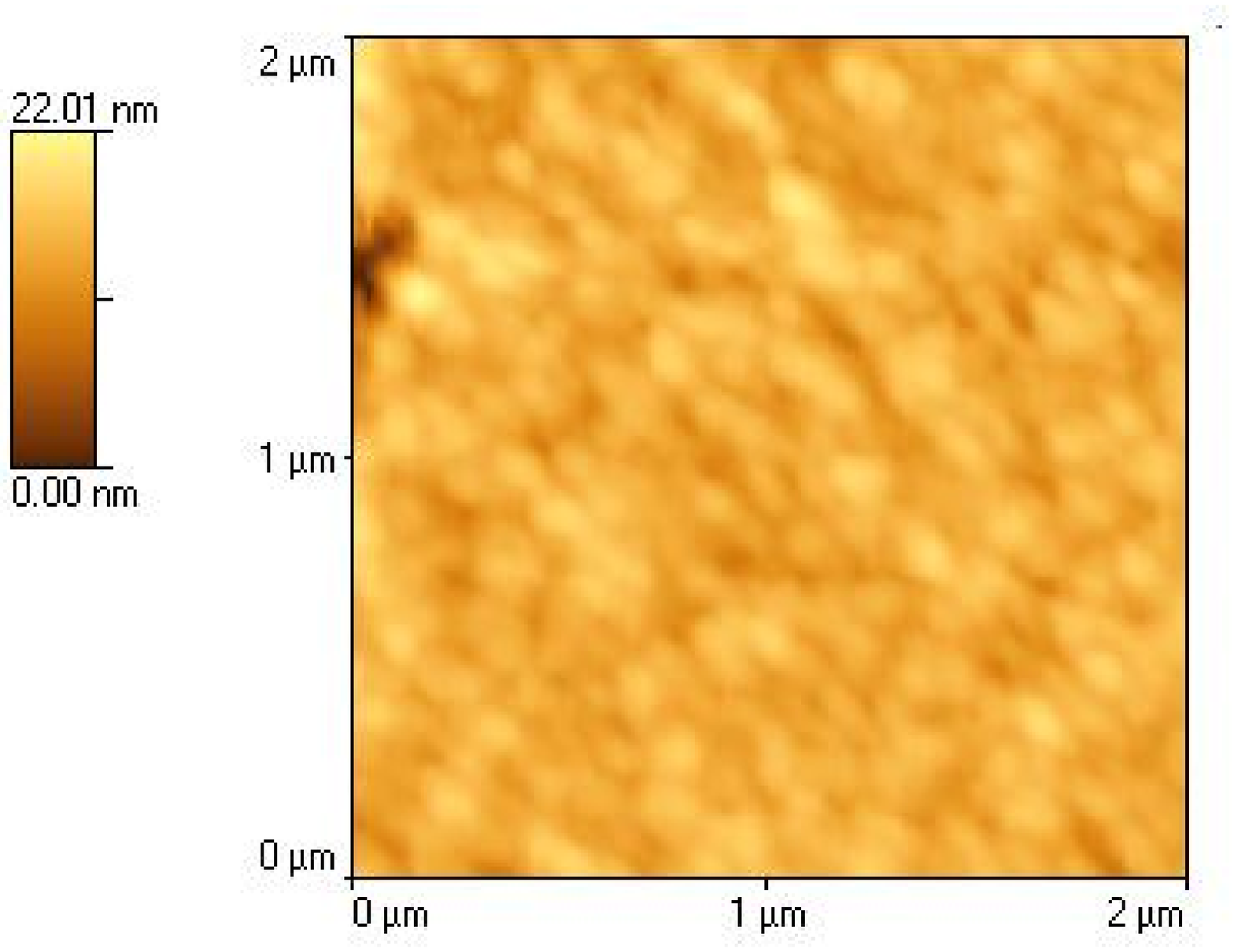, width=2.5in}
\caption{\sl The X-Ray diffractogram of asgrown nanocomposite thin film 
(G1) along with it's AFM image.}
\end{figure}
%%%%%%%%%%%%%%%%%%%%%%%%%%%%%%%%%%%%%%%%%%%%%%%%%%%%%%%%%%%%%%%%%%%%%%%%%%%%%%%%%

\section{Results and Discussions}
Films of various thickness were fabricated and studied. Of all the samples
studied, we report and compare the results of three extreme thicknesses here 
(in accordance of thickness G1$<$G3$<$G2).

\subsection{X-Ray Diffraction \& Morphological Studies}

Asgrown sample of G1 show a lone peak at ${\rm
2\theta=36.895^o}$ (fig~1). JCPDS card (No. 36-1451) reports 
this as the (101) peak of ZnO. This corresponds to the wurtzite structure 
that bulk ZnO is known to crystallize in with lattice constants 
a$\sim$0.3250nm and c$\sim$0.5207nm. The XRD peaks are considerably broadened 
and are
indicative of very small size particles and inturn existence of nanocrystals.
 The average crystallite size of the particles in the film can 
be calculated by Scherrer's formula.

\begin{eqnarray}
D={0.9\lambda \over Bcos \theta}\label{e2}
\end{eqnarray}
%%%%%%%%%%%%%%%%%%%%%%%%%%%%%%%%%%%%%%%%%%%%%%%%%%%%%%%%%%%%%%%%%%%%%%%%%%%%%%%%%
\begin{figure}[h]
\epsfig{file=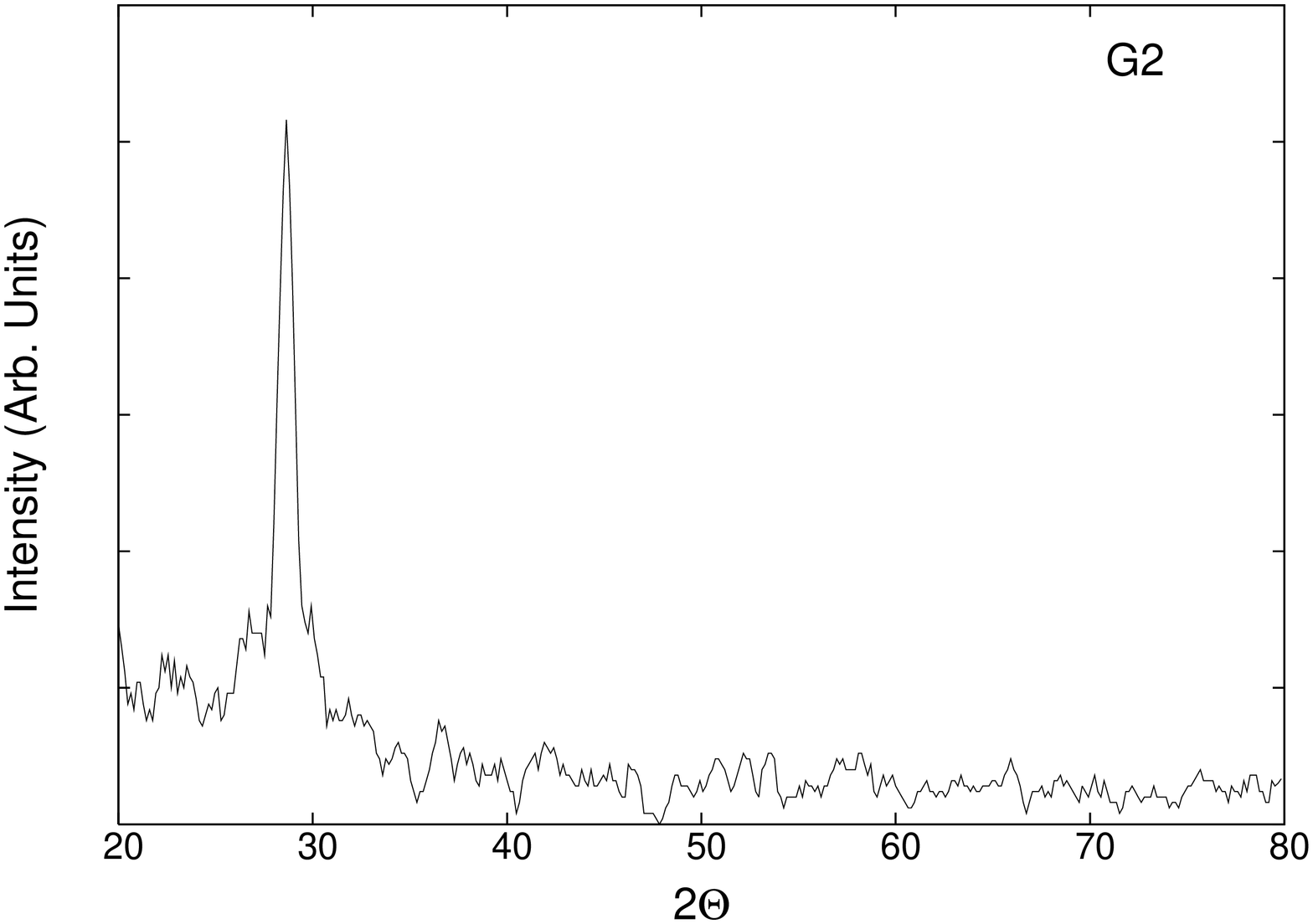, width=2.25in, angle=-90}
\vfil
\vskip 0.3cm
\epsfig{file=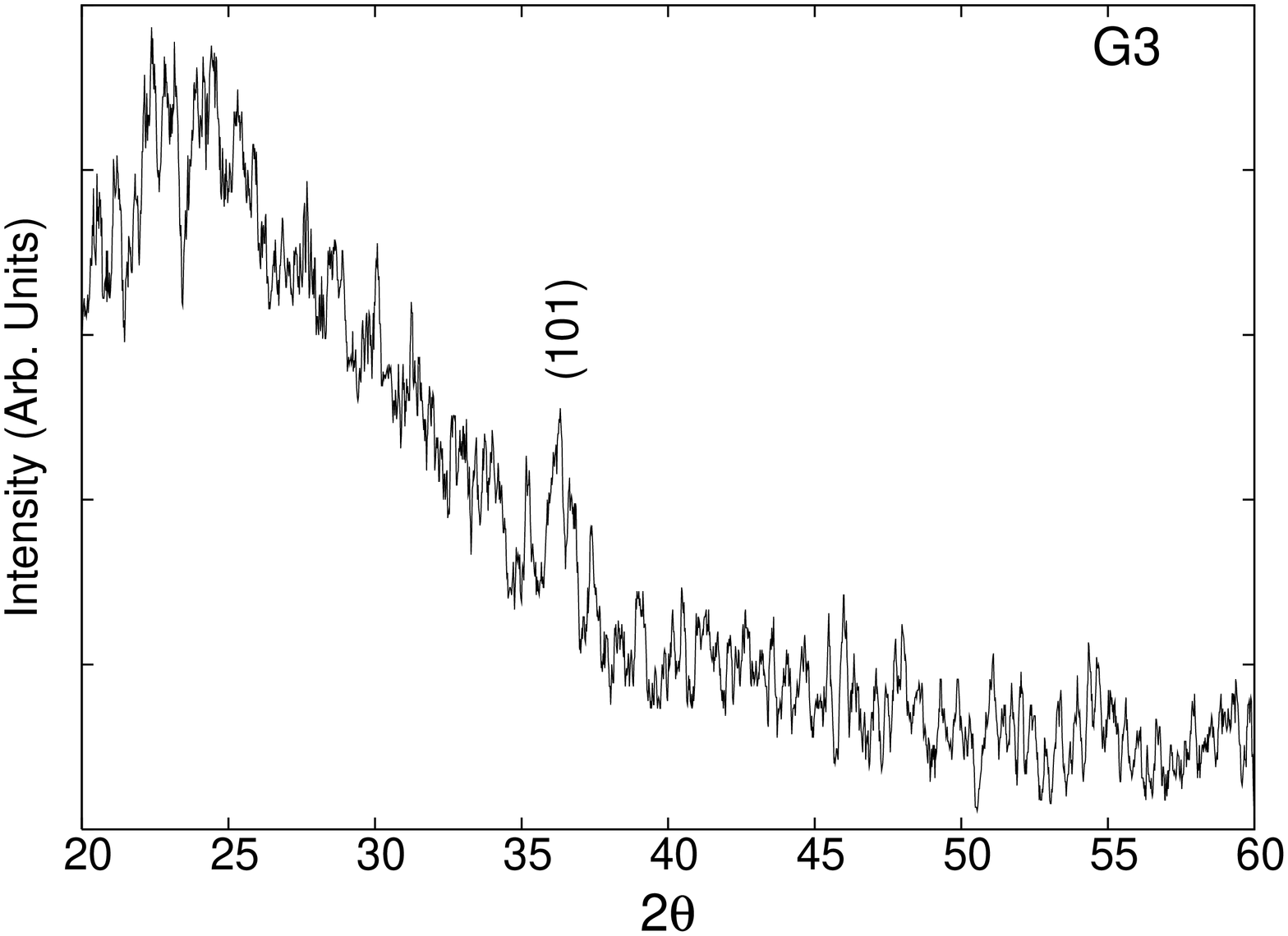, width=2.5in, angle=-90}
\caption{\sl X-Ray diffractograms of asgrown nanocomposite thin films, G2
and G3. On comparing with
the diffractogram of G1 shown in fig~1, it is clear that the lone peak of ZnO
diminishing with increasing film thickness. }
\end{figure}
%%%%%%%%%%%%%%%%%%%%%%%%%%%%%%%%%%%%%%%%%%%%%%%%%%%%%%%%%%%%%%%%%%%%%%%%%%%%%%%%%

where $D$ is the grain size (in $\AA$), $B$ is the FWHM of the particular
peak in radians, ${\rm \theta}$ is the Bragg's angle and $\lambda$ ($\sim
1.5405\AA$) is the wave-length of X-ray. The size of the grains in the 
asgrown sample of G1 was computed to be 20nm. 
The hump seen in the X-ray diffractogram 
show the short range ordering of silicon atoms in the background. Such hump
has been observed in amorphous silicon studied by Gwo-Mei et al \cite{gwo}.
While the X-Ray diffraction pattern  
prove the existence of nanocrystallinity, a direct verification of their 
existence and the particle size can be done by examining them with electron 
microscope (results not shown here) or for more localised areas using the 
Atomic Force Microscope (AFM). The average grain size as measured from the
AFM image was ${\rm \sim}$100nm (fig~1). The disagreement between the results
from the diffractograms and AFM images implies that the diffraction peak's
broadening is not just due to the grain sizes. X-Ray peaks are known to
become broad due to the existence of stress/ defects in the crystal.
\cite{bada} Thus, we expect the nanocrystals of ZnO formed in the film to
be in a stressed state with defects in the crystal structure. Such
observations of stress in ZnO films grown on silicon substrates have been
reported and explained on the basis of large mismatch between the two
material's lattice constants.\cite{for13} The stressed condition are
expected to manifest itself in the optical properties of the samples.

The asgrown films of G2 too show a lone peak, however at ${\rm
2\theta=28.485^o}$. This is the characterisitc (111) peak of silicon. The 
average size of silicon grains in our G2 asgrown sample as calculated using
the Scherrer's formula was found to be 10nm. It can be noticed that as
the film thickness increases, the nanocrystallinity of ZnO is screened by
the increasing ordering of silicon (fig~2). This is unlike in G1 where a
well resolved peak of ZnO stands out compared to the short range ordering of
silicon. With increasing thickness of the film the ordering of silicon
atoms improves and the hump corresponding to short range order gives way 
to a peak. Peaks of Zinc oxide are not evident, however nanocrystals of ZnO
can not be ruled out. It was not possible to resolve whether two distinct 
morphologies exist from
the AFM images (not included). Evidence for the coexistence of two 
nanocrystals however was resolved from our optical characterisations.  

\subsection{Optical Studies}
%%%%%%%%%%%%%%%%%%%%%%%%%%%%%%%%%%%%%%%%%%%%%%%%%%%%%%%%%%%%%%%%%%%%%%%%%%%%%%%%%
\begin{figure}[h]
\epsfig{file=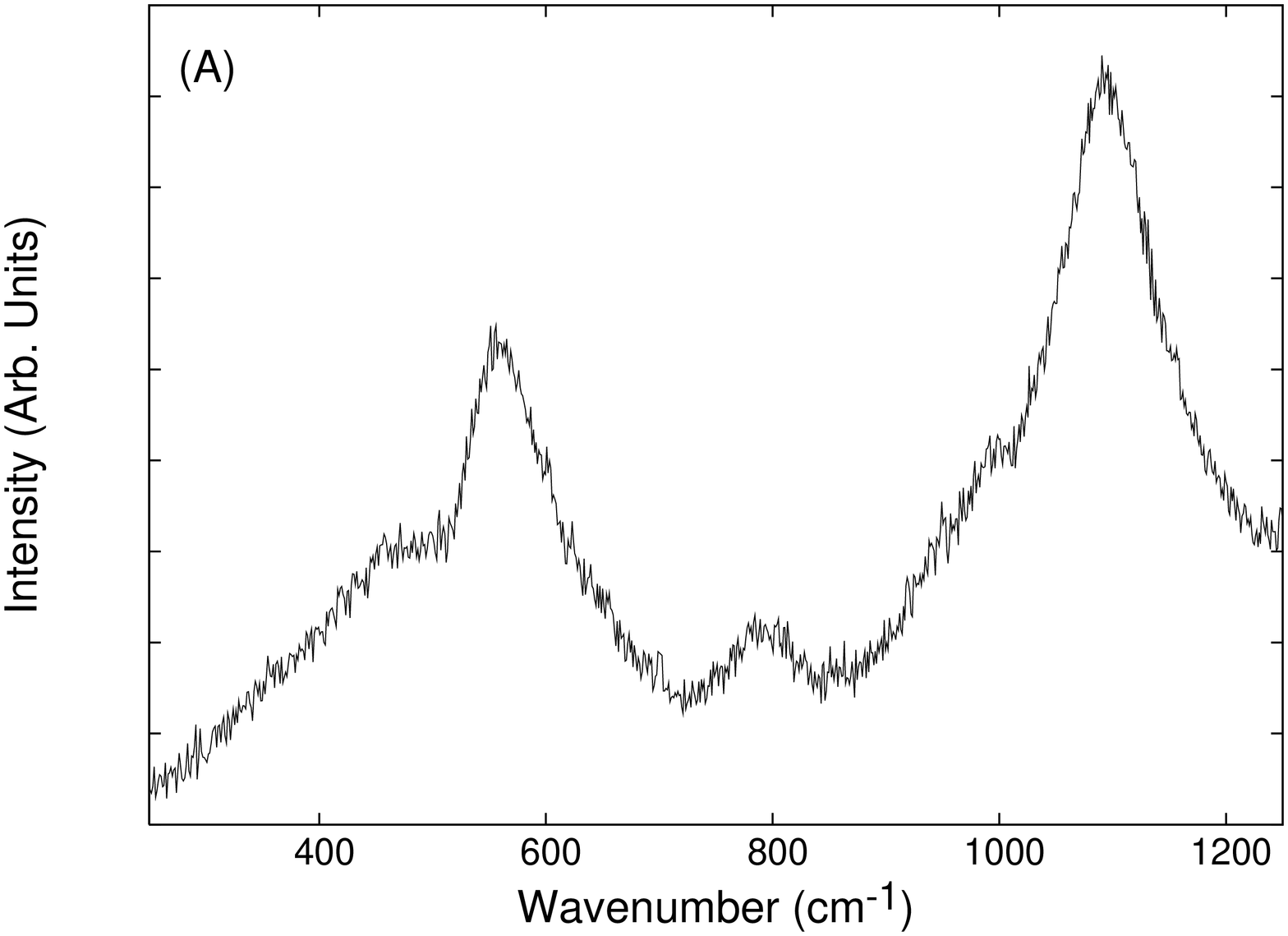, width=2.3in, angle=-90}
\hfil
\epsfig{file=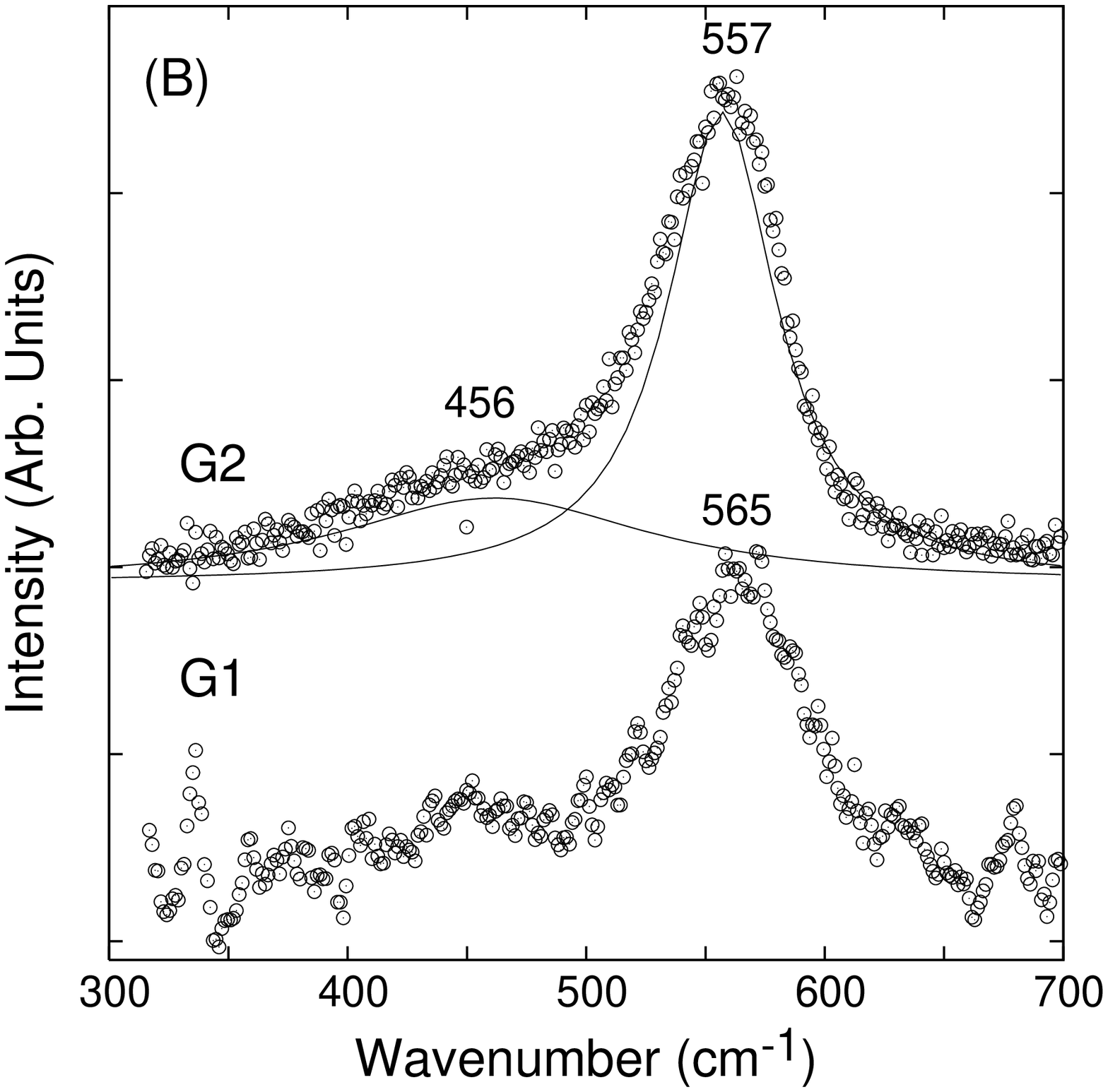, width=2.5in, angle=-90}
%\vskip -0.5cm
\caption{\sl The Raman spectra of (A) G1 sample and (B) compares spectra of
G1 and G2 sample in range 300-700${\rm cm^{-1}}$.}
\end{figure}
%%%%%%%%%%%%%%%%%%%%%%%%%%%%%%%%%%%%%%%%%%%%%%%%%%%%%%%%%%%%%%%%%%%%%%%%%%%%%%%%%

Raman scattering was performed to investigate the vibrational
properties of the ZnO nanocrystals and Si in which the nanocrystals are
embedded. The Raman spectra was taken in standard back scattering geometry 
with ${\rm Ag^{2+}}$ laser used as an excitation source. Vibrational spectra 
from ZnO nanocrystals, silicon background and the glass substrate can be 
expected. However, silicon oxide has no contribution to Raman 
spectrum.\cite{PhysE} Hence contributions from the glass substrate can be ruled 
out. The Raman spectra of ZnO has been well documented and reported to have 
both longitudinal optical (LO) and transverse optical (TO) phonon frequencies 
that split into two frequencies with symmetries ${\rm A_1}$ and 
${\rm E_1}$.\cite{rr19} Prominent among the peaks is the ${\rm A_1(LO)}$ mode 
vibration 
of ZnO at ${\rm 570cm^{-1}}$. Other important Raman peaks of zinc oxide are 
the ${\rm E_2}$ intense peak at ${\rm 438cm^{-1}}$ and the weak 
${\rm E_1(LO)}$ peak at ${\rm 
585cm^{-1}}$. Since this peak (${\rm 585cm^{-1}}$) is very close to the 
${\rm A_1(LO)}$ peak (${\rm 570cm^{-1}}$), both usually result in an
apparent single broad peak. Along with these major peaks, two weak peaks at 
${\rm 332cm^{-1}}$ and ${\rm 381cm^{-1}}$ have been reported.\cite{r23}

\par Besides these vibrational peaks, a prominent LO mode is expected around
1100-1150${\rm cm^{-1}}$.\cite{ram10, ram11, ram12, ram13} The position of
this peak in G1 is at 1095${\rm cm^{-1}}$ (fig~3A). Peak width and shifts 
indicates the size of the nano-crystals. As the crystallite size 
increases the peaks becomes broader and shifts to lower energy.\cite{size3,
size4, size5, size6} Thus our films contain small sized 
nanoparticles of ZnO. Other observed peak 
associated with zinc oxide was the ${\rm E_1(LO)}$ mode at ${\rm
585cm^{-1}}$. However, prominent by
it's omission is the ${\rm E_2}$ peak around ${\rm 438cm^{-1}}$. This peak
is associated with the wurtzite crystal structure of zinc oxide. The absence 
of intense Raman-active mode at ${\rm 438cm^{-1}}$ and 
existence of peak at ${\rm 585cm^{-1}}$ in our samples 
corroborate our X-Ray diffraction results that our samples have possible 
structural 
defects like oxygen vacancies, zinc interstitials and free
carriers etc.\cite{r12}

Zinc oxide nanocrystals of sample G1 and G3 are embedded in a matrix
background of amorphous silicon. The Raman spectrum of amorphous Silicon
(a-Si) consists of several broad bands, the transverse
acoustic (TA) is found at ${\rm 150cm^{-1}}$ and the longitudinal acoustic band 
(LA) at ${\rm 310cm^{-1}}$.
As expected a sharp peak that can be ascribed to a-Si can be seen ${\rm 
\sim 300cm^{-1}}$. X-Ray diffraction results suggested the 
background of G2 samples to consist nanocrystalline Silicon (n-Si). Evidence
of this was also found in our Raman studies. The spectrum of n-Si is very
similar to that of the single crystal silicon (c-Si) with a peak appearing at 
${\rm 519.4cm^{-1}}$. However, additional peaks appear at 604 and ${\rm
423cm^{-1}}$ in n-Si samples.\cite{ok18} As a result we expect a broadening
and existence of a multi-peak band between 400 and 600${\rm cm^{-1}}$.
Comparing the Raman spectra of G2 and G1 between these two wavenumbers
(fig~3B), we can appreciate the broadening that takes place in sample G2. 
A deconvolution was performed on the Raman spectra of G2 by assuming the 
peaks to have Lorentzian shapes. The peak shown at ${\rm 456cm^{-1}}$
couldn't be deconvoluted further due to the lack of prominent shoulders in
the spectra, but it should be a multi-peak band comprising 
the ${\rm 423cm^{-1}}$ peak of n-Si and ${\rm 438cm^{-1}}$ peak of ZnO.
The confidence with which deconvolution can be done using standard softwares 
like Origin6.0 depends on the shoulders present in unresolved
peaks. Since, we did not have many shoulders, we did not try to force more
than two peaks for deconvolution.
%%%%%%%%%%%%%%%%%%%%%%%%%%%%%%%%%%%%%%%%%%%%%%%%%%%%%%%%%%%%%%%%%%%%%%%%%%%%%%%%%
\begin{figure}[h]
\epsfig{file=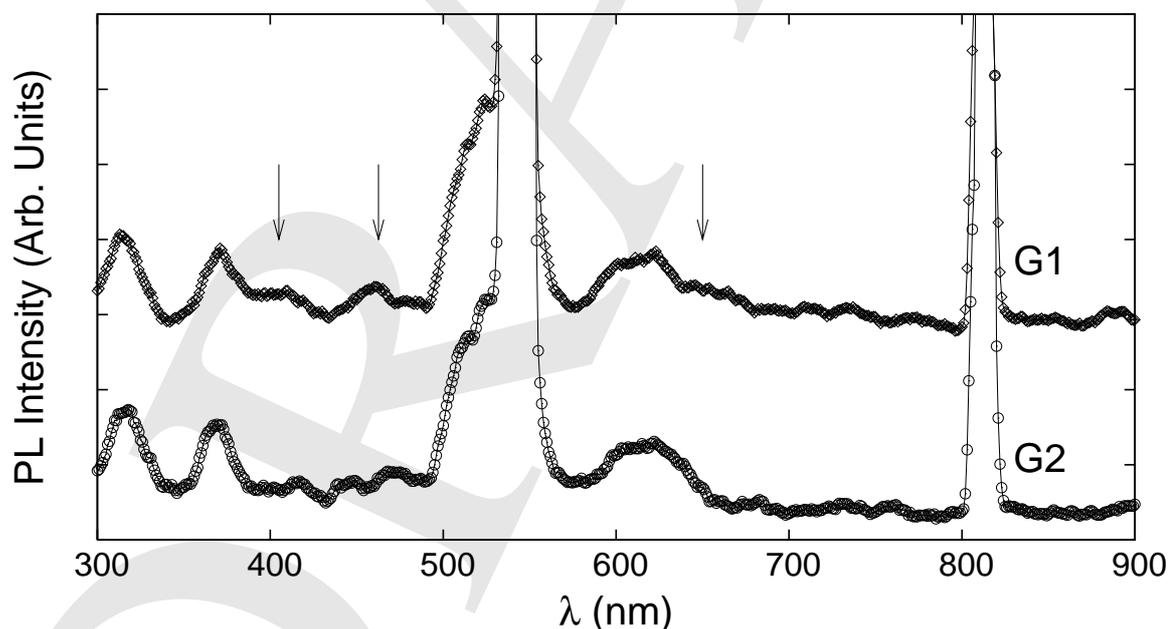, width=3.25in, angle=-90}
\caption{\sl The PL spectra of G1 and G2 asgrown samples. The spectras are
near identical with peak broadening (indicated by arrows) in G1 sample, 
where ZnO nanocrystals are embedded in a-Si.}
\end{figure}
%%%%%%%%%%%%%%%%%%%%%%%%%%%%%%%%%%%%%%%%%%%%%%%%%%%%%%%%%%%%%%%%%%%%%%%%%%%%%%%%%

The results from Raman spectra analysis reinforces our conclusions made from
structural and morphological studies and confirm the formation of ZnO quantum
dots in a matrix amorphous background of silicon in samples G1 and G3. 
In sample
G2, the quantum dots of ZnO are suspended in a matrix of nanoparticles of 
silicon. Zinc oxide in the nano regime exhibits strong photoluminescence, 
hence a direct proof of
the existence of ZnO in nanocrystalline state can be obtained from the PL 
peaks. As stated above, the peak at ${\rm 570cm^{-1}}$ in Raman spectra 
originates due to structural defects (oxygen vacancies, zinc interstitials 
and/or free carriers etc.) and impurities.\cite{r12} It has been reported
that such structural defects in ZnO nanocrystals give rise to new peaks in
PL spectra. 
%%%%%%%%%%%%%%%%%%%%%%%%%%%%%%%%%%%%%%%%%%%%%%%%%%%%%%%%%%%%%%%%%%%%%%%%%%%%%%%%%
\begin{figure}[h]
\epsfig{file=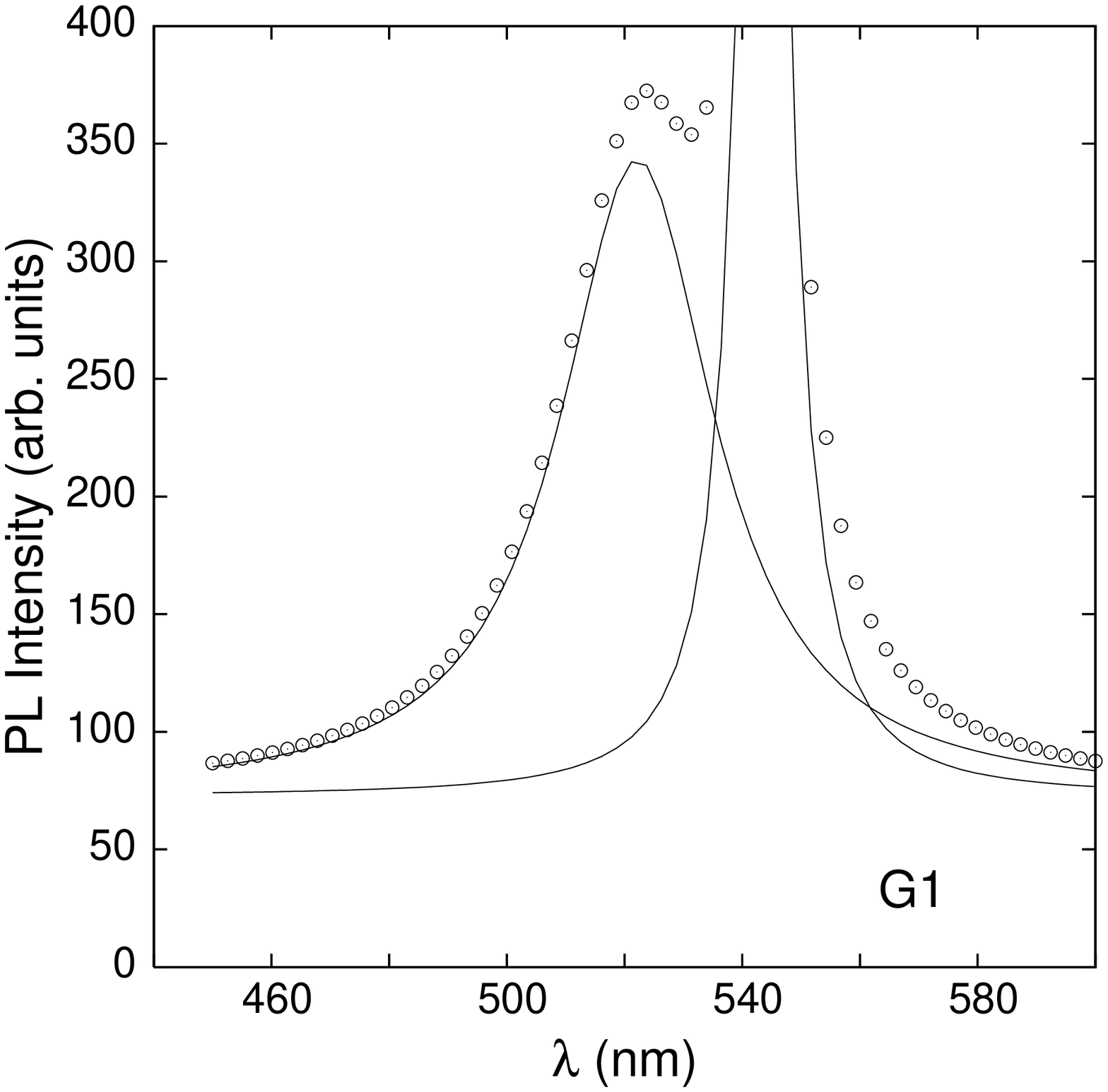, width=2.75in, angle=-90}
%\vfil
\epsfig{file=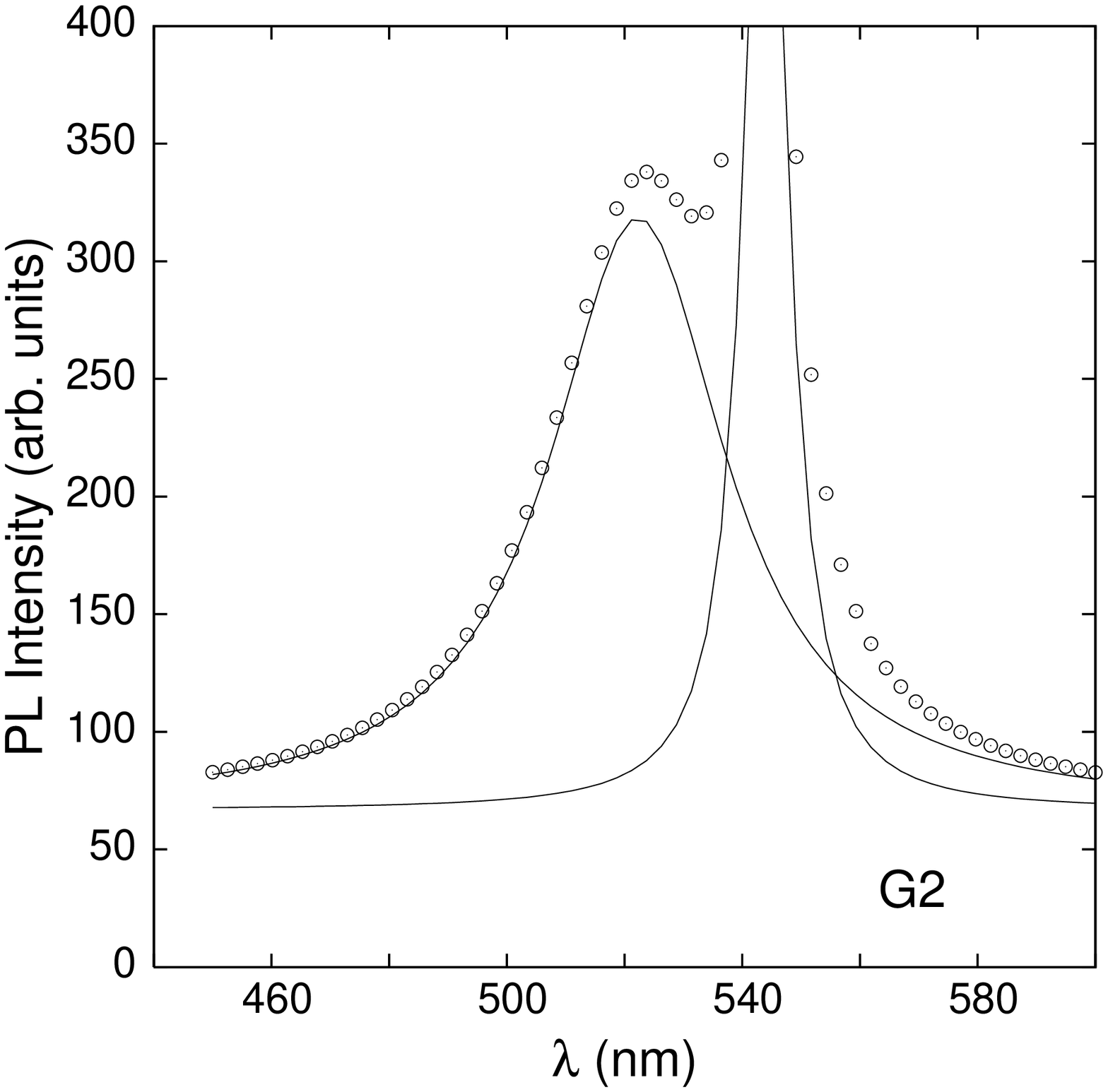, width=2.75in, angle=-90}
\caption{\sl The green band in PL spectra was deconvoluted to resolve the
emission peak from the intense ${\rm 2^{nd}}$ harmonic peak of the 270nm
excitation source.}
\end{figure}
%%%%%%%%%%%%%%%%%%%%%%%%%%%%%%%%%%%%%%%%%%%%%%%%%%%%%%%%%%%%%%%%%%%%%%%%%%%%%%%%%

The photoluminescence (PL) spectrum for the G1 and G2 samples are shown in
Fig~4. The emission spectra were obtained using 270nm excitation source.
The spectra were recorded without filters and hence we see two lines, the 
${\rm 2{nd}}$ and ${\rm 3^{rd}}$ harmonics of the source at 540 and 810nm
respectively. Characteristic peaks of ZnO and Si are observed at 310nm, 
365nm, 522nm and 620nm. ZnO nanoparticles giving PL emissions in the UV
region have
been reported\cite{eva, cong} and have been explained due to the radiative
annihilation of excitons. The position of the green emission peak ($\sim$ 
520nm) was resolved by deconvoluting the emission peak and ${\rm 2^{nd}}$ 
harmonic line spectra (see fig~5). The peak position is in good agreement 
with earlier reports.\cite{van, nsc} The 
green emission of ZnO nanocrystals are explained due to transition of
electrons from the conduction band to trapped hole within the band gap.
Oxygen vacancies, in other words defects, are considered to be the main
candidate for recombination centers involved in this emission.\cite{dj1, dj2} 
Our XRD analysis with peak broadening suggested that our samples had
defects. Defects were also indicated by our Raman spectra. This is confirmed
by the green emission band in PL studies. The 620nm emission spectra belongs
to the silicon matrix of our samples.\cite{bmr}   

The PL emission spectra (fig~4) of G1 and G2 samples appear identical. While
G1 sample had ZnO nanocrystals embedded in a matrix of amorphous silicon, 
ZnO nanocrystals of G2 sample are embedded in nanocrystalline silicon. The 
only variation due to the structurally different matrix (background) seems
to be the disapppears of shoulders in peaks (indicated by pointers in fig~4)
as silicon goes from amorphous to nano-crystalline state. While
this geometry has not given multiple emission peaks in visible region
(unlike case reported where ZnO was grown on porous-silicon
substrate\cite{forun}) it
has given stability to the ZnO nanocrystals. The PL spectra has maintained
its nature over 90 days since sample fabrication. Also, the strength of
emission in our samples is far stronger than that observed in samples of Singh
et al.\cite{forun} We believe the thermal evaporation method used for sample 
fabrication leads to the nanocrystals of ZnO to be homogenously 
envoloped by silicon, thus contributing to the strong emission
spectra. Thus, this method of fabrication might result in
more efficient opto-electronic devices designed around the feature of 
green emission.

\par If the emission peak due to silicon at 620nm can be made strong and
intense, it would merge with the peak due to ZnO at 520nm and result in a
broad-band spectra. We believe that strong broad band emission can be 
obtained by the simple fabrication method we have adopted. In order to 
obtain this, the parameters such as film 
thickness, substrate temperature, amount of ZnO dispersed in Si matrix and 
the other deposition conditions have to be optimized. Such studies are being
done by our group. 

\section{Conclusion}
ZnO:Si nanocomposites were fabricated by co-thermal evaporation of ZnO and
Si powder. Resulting films yielded nanocrystalline ZnO embedded in amorphous
Silicon or nano-crystalline Silicon, depending on the film thickness. The
PL emission spectra of these films present two neighboring emission peaks at 
520 and 620nm due to
ZnO nanocrystals and silicon respectively. On optimization, broad-band
emission could eventually result white light emission. The method holds promise as
potential low-cost fabrication method for white LEDs etc.

\section*{Acknowledgment}
The resources utilized at University Science and Instrumentation Center, 
University of Delhi and Geology Department, University of Delhi is gratefully 
acknowledged. Also, the resources used at the Instrumentation Center and
University Information Resource Center, Guru Gobind Singh Indraprasta
University is also acknowledged. We also would like to express our gratitude to Mr. Kamal
Saran (National Physical Lab., Delhi) for carrying out the photoluminescence
studies.

\end{document}